\documentclass[12]{iopart}
\usepackage{graphicx}
\usepackage{color}
\pdfoutput=1

\begin{document}


\title[Locating and observing optical counterparts to gravitational wave bursts]{LOOC UP: Locating and observing optical counterparts to gravitational wave bursts}

\author{Jonah Kanner$^1$, Tracy L. Huard$^2$, Szabolcs M\'{a}rka$^3$, David C. Murphy$^4$, Jennifer Piscionere$^3$, Molly Reed$^5$, and Peter Shawhan$^1$ }
\address{$^1$Physics Department, Univ. of Maryland, College Park, MD 20742, USA

$^2$Department of Astronomy, Univ. of Maryland, College Park, MD 20742, USA

 $^3$Department of Physics, Columbia Univ., New York, NY 10027, USA

 $^4$Observatories of the Carnegie Institution of Washington, Pasadena, California 91101, USA

 $^5$Montgomery Blair High School, Silver Spring, MD 20901, USA}
\ead{jkanner@umd.edu}

\begin{abstract}
Gravitational wave (GW) bursts (short duration signals) are expected to be associated with highly energetic astrophysical processes.  With such high energies present, it is likely these astrophysical events will have signatures in the EM spectrum as well as in gravitational radiation.  We have initiated a program, ``Locating and Observing Optical Counterparts to Unmodeled Pulses in Gravitational Waves'' (LOOC UP) to promptly search for counterparts to GW burst candidates.  The proposed method analyzes near real-time data from the LIGO-Virgo network, and then uses a telescope network to seek optical-transient counterparts to candidate GW signals.  We carried out a pilot study using S5/VSR1 data from the LIGO-Virgo network to develop methods and software tools for such a search.  We will present the method, with an emphasis on the potential for such a search to be carried out during the next science run of LIGO and Virgo, expected to begin in 2009.

\end {abstract}
\pacs{95.55.Ym, 95.75.Wx}
\maketitle{}

\section{Introduction}

A gravitational wave (GW) burst would likely originate with an extremely energetic astrophysical process.  Such an event is likely to emit a significant amount of energy in the electro-magnetic (EM) spectrum \cite{fox_na, syl, sngw}, which is expected to arrive in near coincidence with the GW emission (See however \cite{Kahya}).  Past GW searches have exploited this expected association between gravitational
waves and EM astronomy.  Use of astronomical information on such phenomena as gamma-ray bursts (GRB's) and soft gamma-ray repeaters has lead to a variety of
targeted or ``triggered'' GW searches with sensitivities better than their all-sky, all-time counterparts \cite{bars, 2005_extrig, virgo_extrig, 070201,s4grb,sgr}.  While the details of these searches vary, they all
share a common theme:  available EM information triggers careful searches in the gravitational wave data.

Here, we propose to combine EM and GW data in the \emph{other direction}, by using gravitational wave data to trigger EM observations.  In some sense, this is the
``natural'' way to perform a combined EM/GW search for rare events, since GW detectors are, by nature, sensitive to most of the sky, where most current EM observatories must be pointed.  However, there exist a number of technical and logistical difficulties associated with such a search.

We have begun a program to address these obstacles and perform an interferometer-triggered search for electro-magnetic transients, ``Locating and Observing Optical Counterparts to Unmodeled Pulses in GW'' (LOOC UP) \cite{LOOC}.  A partially-dedicated network of robotic telescopes could, during the next science run of the LIGO-Virgo network, perform a sustained hunt for EM transients associated with low-threshold gravitational-wave candidates.  Such a search would be essentially all-sky and all-time, but would still benefit from the increased sensitivity of a combined GW/EM search.

\section{Motivation}

There are several motivations for performing targeted transient searches using GW triggers.  Most exciting is the possibility of hastening the first confirmed
detection of a GW signal.  The noise in interferometers is non-stationary and contains transients \cite{allsky}. If a true GW signal were present in the data with a relatively low signal to noise ratio (perhaps SNR of 5 to 10 in each detector), it could be difficult or impossible to make a strong case for an astrophysical, rather than environmental, origin.  However, association of such an event with an astrophysical transient could drastically alter the situation.  By actively seeking the optical transients to low-threshold events, we can effectively increase the sensitivity of our GW search.  The sensitivity of the Enhanced LIGO/Virgo+ science run (labeled S6/VSR2, to start in 2009) \cite{enhligo,advirgo} is expected to be at the threshold of making detections likely \cite{richard}.  Pushing the interferometer range during this time will be extremely important.

Even a high SNR GW detection would benefit from an astronomical confirmation.  An associated EM signal would erase any doubts about the astrophysical origin of the GW signal.  Further, the measurement of an EM counterpart would carry important information about the GW source mechanism.  For example, detection of an optical signal would greatly enhance the precision of the source localization, potentially allowing the identification of a host galaxy and associated red shift.  Further astrophysical uses of EM data are potentially endless, as is evidenced by the wealth of progress based on gamma-ray burst afterglow observations \cite{shb_rev}.

Another motivation stems from the current interest in gamma ray bursts.  A GW triggered transient search is effectively a targeted search for GRB afterglows.  Short-hard GRB's may emerge from compact object mergers, which also emit significant gravitational radiation \cite{fox_na}.  A merger event with the gamma rays beamed away from earth (off-axis) would be invisible to gamma-ray surveys, such as SWIFT \cite{swift}.  However, the gravitational waves, as well as the afterglow radiation, would be emitted in a wider range of directions.  Thus, a GW triggered search could hope to find the ``orphan'' afterglow for such an off-axis event.  Early indications suggest that short-hard bursts have beaming fractions of one to a few percent, or opening angles of $\sim$10 degrees \cite{fox_na}.  This suggests that off-axis events are far more common than on-axis events, i.e. in a fixed volume of space, the rate of orphan afterglows could be significantly greater than the rate of observable GRB's.  This motivates the GW triggered search as complimentary to satellite based searches for the rarer, \emph{on-axis} GRB's.

It may be noted here that an association between a GW event and an optical transient may also be made by using the data from untriggered optical transient hunts \cite{OSU, stubbs}.  Such searches scan the sky, and during the S6 era, are expected to cover large regions with repeat times of $\sim$4 days \cite{stubbs}.  Since these searches could make an association between GW and EM signals with a very different time-scale than a triggered observation (4 days vs. minutes to hours), a search of this data set could serve as an excellent complement to the LOOC UP project.

The GW community is working towards an era where reports of GW events will be commonplace.  Significant upgrades in detector hardware will usher in the advanced detector era, when Advanced LIGO and Advanced Virgo anticipate observing multiple bursts per year \cite{advirgo,advligo}.  During this time, GW observations of supernovas and other events will be most useful to other observatories if they are reported promptly.  Given that this vision of joint GW/EM astronomy may be fully realized as early as 2014, the time to begin preparing the computing infrastructure and software for real-time GW data analysis is now.  We should count real-time analysis software among the many new advanced detector era technologies being developed for S6/VSR2.

\section{Source Models}

It is difficult to estimate the exact nature of the EM counterpart to a given GW burst event.  For example, while the GW energies of compact merger events may be calculated
very precisely, predicting the strength of the coupling to the EM channel is a dubious undertaking.  However, several rough estimates and models have guided our thinking in constructing our proposed search.

Li and Paczy\'{n}ski have proposed that, during a double neutron star (DNS) merger or a black hole/neutron star merger, some fraction of the neutron star mass may be ejected \cite{li}.  Some of this ejected, neutron rich matter would then decay, releasing energy and fueling an expanding, glowing fireball.  The luminosity of such an event peaks after about a day, and then falls off over the course of a few days.  At 40 Mpc, the apparent R band (red filtered) magnitude of this nuclear fireball is estimated at $\sim$15 \cite{syl}.

Empirical evidence for our sought counterparts exists in the form of observed afterglows of short-hard GRB's.  To imagine how such an event would appear, we can take short GRB 050724 as an example.  Its afterglow was identified with a red-shift of z = 0.26 and an optical flux density of $\sim$0.03 mJy 12 hours after the event \cite{edo, 050724}.  Neglecting extinction, we could expect that the same event, at 40 Mpc, would yield an apparent magnitude of $\sim$13.  That is, GRB afterglows, which are observable at cosmological distances, could appear relatively bright if placed at Enhanced LIGO/Virgo+ distances.  However, this estimate is based on observed \emph{on-axis} GRB afterglows.  It is possible that an off-axis GRB afterglow could only be observed at lower flux due to beaming effects.

Finally, we note that supernovas are also potentially sources of both GW and EM emissions.  While estimates of GW emissions from supernovas are uncertain, they are generally thought to be less energetic in GW's than compact object mergers \cite{sngw}.  If we place a supernova at 40 Mpc, the apparent magnitude at peak luminosity would be between 14 and 17 \cite{snem}.  It is likely that a supernova would have to be much closer than 40 Mpc to be observable in GW during S6/VSR2.  So, any supernovas within interferometer reach should have EM counterparts with apparent magnitude $\leq$ 15.

\section{Search Overview}

\begin{figure}
  \scalebox{0.25}{\includegraphics* [width=40cm]{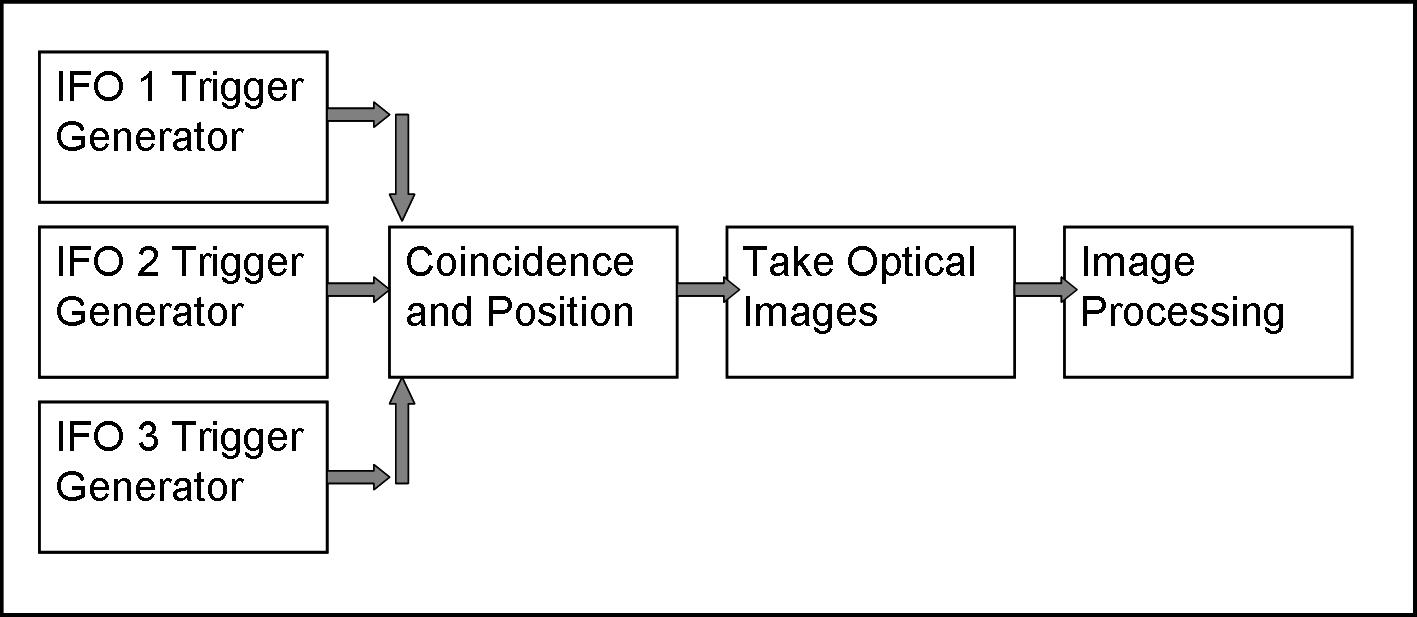}}\\
  \caption{A schematic of the analysis.  Triggers are identified in data from the three sites of the LIGO-Virgo network.  The three trigger lists are then compared to find coincident events, labeled ``candidate events.''  A sky region is assigned to each event candidate.  The sky region is then imaged with an EM observatory. These images are searched for transients.}\label{Figure 1}
\end{figure}

With the above models as guides, we can make the following outline for a search (see figure \ref{Figure 1}).

The process begins with GW data analysis.  Short time intervals of GW data are Fourier transformed, and regions of the time-frequency plane with excess power are
identified as ``triggers.''  A number of software packages exist for GW burst trigger generation, differing in a variety of details \cite{kw, qs, bn}.  When triggers appear simultaneously (within light travel time) in all three sites of the LIGO-Virgo network, they are identified as a candidate event.  The requirement of triple coincidence is important for two reasons.  The first is common to nearly all sensitive GW searches: some level of coincidence is necessary to reduce the rate of spurious events \cite{allsky}.  Additionally, since GW burst source position estimation relies primarily on time delays between interferometers, 3 site coincidence is required to obtain a reasonably localized candidate.

After identifying an event candidate, an estimate of the source position may be calculated.  The precision of a GW source position estimation depends on the choice of algorithm and the strength of the GW signal.  With three detectors and a moderate signal it can likely be accomplished to within a few degrees \cite{cwb,inc,incplus}.  We will consider different position reconstruction algorithms in \sref{comp}.

The location information is then used to create an observing schedule for some telescope or telescope network.  In the title of this paper, we focus on optical counterparts; however, infrared, radio, and x-ray are all equally valid bands to search.

Within about 1 hour of the GW detector event, the first images are taken of the estimated sky position.  Follow-up images should occur on timescales of a few hours and a few days, both as references to identify variability and as time-domain traces of the light curves of any potential transients.  Given the relatively bright nature of our targeted sources, imaging to a magnitude of $\sim$15 should be sufficient.

Finally, images are reduced, cataloged, and searched for transients.  Creating a real-time pipeline for image processing would allow the quick identification or rejection of transients from images, helping to guide the observing schedule.

\section{Pilot Study} \label{pilot}

During the summer of 2007, we performed pilot studies using current data from the LIGO-Virgo network to develop our analysis tools and observing strategy.  We successfully identified event candidates in real-time, estimated source locations, and imaged selected areas of the sky.

\subsection{Galaxy Catalog}

One difficulty with a GW triggered transient hunt is the relatively poor localization of the source with interferometer data alone.  A GW position estimate might have an area of $\sim$10 square degrees.  This is larger than the field of view of most astronomical instruments, and so a single image may not capture the full estimated source location region.

We have developed a technique to overcome this localization problem.  The method takes into account the limited reach of the interferometers.  Given some maximum range, we can construct a finite catalog of plausible host galaxies.  When a candidate event is found, we may search our catalog for hosts that are consistent with both the position reconstruction and the interferometer range.  In this way, we may image the area immediately around likely host galaxies, rather than requiring that our telescope network image the entire GW error ellipse.

\begin{figure}
  \includegraphics*[width=12cm]{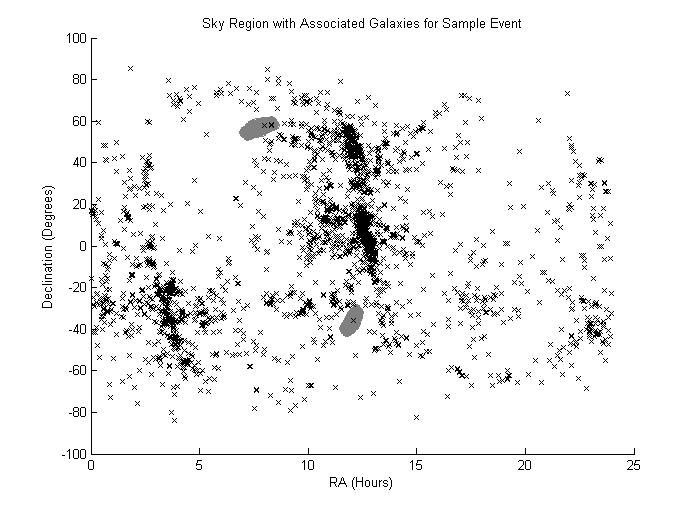}\\
  \caption{The location of Milky Way globular clusters and galaxies within 20 Mpc are marked with x's.  The gray areas represents a pair of error ellipses for a sample event candidate with assumed timing uncertainty of 0.5 ms.  The error ellipses contain four potential hosts.}\label{Figure 2}
\end{figure}

For our pilot study during S5, we adopted and modified a catalog of galaxies \cite{cat} created by the LSC Compact Binary Coalescence group.  Based on the position-averaged LIGO DNS range, we cut all objects further than 20 Mpc.  We also applied a selection cut on the mass to distance ratio, and added Milky Way globular clusters.  This led to a catalog with 2766 objects, an average of about .07 objects per square degree.  In \Fref{Figure 2}, the black symbols mark the location of potential hosts in our 20 Mpc catalog.  Also seen is a pair of error ellipses for a sample GW event candidate, with an assumed timing uncertainty of 0.5 ms in each detector.  The four targets that fall in the error ellipse could be imaged for transients.

If we extend the cut-off distance to 40 Mpc, based on the expected S6 position-averaged DNS range, we could expect an average of $\sim$1 target per square degree of sky.  This would allow a search during S6 to be possible, even if wide field of view instruments are not available.

\subsection{Observations}

  We performed observations during three periods in 2007:

1) June 4-6 on the 2.4 m Hiltner telescope at the MDM Observatory at Kitt Peak, AZ \cite{mdm}

2) July 22 - August 1 on the 1 m Swope telescope at Cerro Las Campanas in Chile, operated by the Carnegie Observatories \cite{swope}

3) September 4 - 9 on the 1.3 m McGraw-Hill telescope at the MDM Observatory

During the June observing run, we used GW triggers produced by the KleineWelle software \cite{kw}.  This trigger generator runs online, and we were able to download trigger lists from the 4 km Hanford detector and the Livingston detector with a lag time of about 30 minutes.  Using only the two LIGO sites, we were able to reconstruct a ``sky ring'' of possible source locations. 

For our second and third observing runs, we replaced the KleineWelle trigger lists with Q-online trigger lists \cite{qs}.  Q-online included trigger production for the Virgo interferometer, extending our network to three sites, but had a larger lag time (about an hour).  Using at least three sites seems to be a necessity --- the sky rings from a two site network are simply too large to observe in a reasonable span of time.  Our software would regularly download trigger lists, identify triple coincident events, estimate the source locations, cross-reference with our galaxy catalog, and post a list of potential observing targets to a web site.  The position estimation was done with a
simple time-of-flight algorithm that used only the measured peak times of each trigger.

From the lists of potential targets, we manually created observing schedules ``on the fly.''  We would observe a target at the earliest possibility, and then capture follow-up images hours and days later.  We took images using both near infra-red (I) and red (R) filters.  Over our three runs, we observed a total of 95 targets.  Image processing and data analysis are currently in progress.

\section{Considerations for future searches}

\subsection{Computational methods}\label{comp}

Most searches for gravitational waves in interferometer data are performed off-line, well after the time when the data was collected; attempting to identify
candidate GW events and source position estimates in real time is a somewhat novel approach.  While several studies have explored GW position reconstruction, no other group has actually used GW data to point an instrument.  So, there is a great deal of work to be done in terms of optimizing source reconstruction, both in terms of optimizing the precision of the reconstruction and optimizing the time used to calculate the reconstruction.

Approaches to estimating the source position can be divided into two broad categories:  incoherent and coherent.  In an incoherent approach, data from each interferometer is used \emph{separately} to calculate trigger properties:  particularly, a peak time of the signal in each interferometer.  Trigger lists from each interferometer are then combined to identify coincident events.  The estimation of source position, in an incoherent approach, uses only the trigger properties, and does not directly combine data from separate interferometers.

From a logistics standpoint, this means that triggers can be calculated at the site of each detector, and then only simple text files listing trigger properties need be brought to a single site.  The position estimation calculation is necessarily computationally cheap, since each interferometer contributes only a few numbers to the calculation.  In our 2007 pilot studies, we were able to use incoherent methods to obtain source position estimates in as fast as 45 minutes, limited mainly by the latency of the initial trigger generation.  There are no fundamental barriers to accelerating this approach, and LSC members are currently discussing implementations that would reduce this time to one minute or less.

On the other hand, these relatively straightforward approaches may have some limitations.  Using only the peak time information from each trigger leads to a solution of 2 distinct sky regions with a three detector network.  If only timing information is available, no method will resolve this degeneracy.

In addition, an incoherent approach may have errors associated with signals containing energy in both polarization orientations.  Time-of-flight calculations assume that the peak times registered in two separate interferometers correspond to the same phase front.  This is not guaranteed for non-aligned detectors --- different detectors are sensitive to different polarizations, and so may observe the peak at different points in the phase.  However, this effect is somewhat mitigated in that only certain regions of the sky are susceptible, as discussed in \cite{incplus}.

While the effect may be limited to some fraction of signals, the worst case scenarios add a significant error to the timing-only solution.  To gain a feeling for the magnitude of the error, we can imagine 90 degrees of phase shift in the measured peak times between LIGO Livingston and Virgo on a 100 Hz signal.  This implies 2.5 ms in timing error on a 26 ms baseline, leading to between 5 and 10 degrees error for most sky directions.  Such an error would make followup observations difficult for all but the largest field-of-view instruments.

Coherent methods have the potential to address the accuracy limitations of an incoherent method, at the expense of introducing logistical and computational issues.
Coherent methods combine sampled time series data, not simple trigger properties, from different detectors before estimating a position \cite{cwb,xp}.  The appeal of such methods is that they seek a signal reconstruction which best fits the data:  they can consider solutions which contain both GW polarizations, thus eliminating the issue which could hamper an approach assuming linear polarization.  In addition, they naturally incorporate amplitude information as well, and so will tend to break the two-fold degeneracy of timing-only incoherent solutions.

There are two major issues that one encounters when trying to use a coherent method real time:  data transfer and computation time. Neither issue is insoluble, but both must be addressed if we wish to obtain a coherent GW source estimation with under one hour latency.

During S5, data transfer was accomplished through a process of ``publishing'' to central locations, such as California Institute of Technology.  Data was typically available with about a 30 minute lag time, but this time would vary.  For a coherent solution, the calculation begins after data transfer, so reliably meeting the one hour mark demands an improvement on this system.  The amount of data (64 kB/s/detector) is not fundamentally problematic, but creating a simple and reliable transfer system will take effort.

Another issue is the computation time for a coherent source reconstruction.  X-pipeline \cite{xp}, for example, currently uses 1 to 3 hours of CPU time for a single event.  The heavy computational cost arises because a segment of $h(t)$ data must be used at each of thousands of trial sky positions to find the best match.  There are several strategies that one can imagine to improve the speed of this process:

1) The trial sky positions may be more optimally chosen, which can improve computation time by perhaps a factor of 2.

2) The length of the data segment used in the calculation may be decreased.  Currently, X-pipeline uses 2 seconds of data at each sky position, though it typically targets bursts with durations less than 0.1 seconds.  This could decrease computation time by as much as a factor of 10.

3)  The timing-only solution could be used to limit the sky grid.  That is, rather than scan over the whole sky, perform the coherent calculation only for positions that
satisfy the constraints of the incoherent calculation.  A demonstration of a technique which is similar to this in spirit has already been successfully performed \cite{incplus}.

It is, of course, also possible that an entirely different software package may prove effective (such as Coherent Event Display \cite{cwb}), but similar issues would likely need to be addressed.

\subsection{Telescope resources} \label{scopes}
In order to perform a sustained, low-threshold search during the S6/VSR2 science run, we need to identify telescope resources that can be used over an extended period of time.  Nearly any band is appropriate, since afterglows of short-hard GRBs have been observed in radio, optical, and x-ray.

In optical, the demands are fairly modest.  The brightness of our target objects (R band magnitude $\sim$15) means that we do not need to search especially deep, and the use of the galaxy catalog means that we do not need to search an especially large field (perhaps 30 arc minutes for each target).  However, we do require multiple, short duration observations on a nightly basis over a period of about a year, so it would be preferable to have one or more instruments that could be operated remotely.  Having access to different instruments on different parts of the globe would expand our coverage, both in the sense of being able to observe targets promptly and to gain greater coverage of the sky.  It is not necessary that any facility commit all its time to this project - a commitment to a few observations a night for about a year would allow for a novel, low-threshold gravitational wave search.

In our pilot study, the devices we used were not exceptional by today's research standards.  However, they met our modest imaging needs.  One possibility for our search is to identify one or more such 1-meter scale telescopes that could be used with minimal expense to other astronomical research fields.

Existing survey projects might be interested in collaborating on this search.  Robotic telescopes used in projects such as ROTSE \cite{rotse} and RAPTOR \cite{raptor} boast large fields of view and fast slewing times; they would be ideal.  With such powerful hardware, making a few interferometer triggered observations a night could be a small time investment towards a study with a potentially large scientific pay-off.  Further, a targeted search for GRB orphan afterglows is consistent with the scientific goals of such projects.

We have also considered purchasing half-meter scale ``off-the-shelf'' telescopes.  This possibility is intriguing, but would demand a substantial effort to set in place all the necessary hardware.  Radio astronomy is also an exciting possibility.  Facilities exist around the world, and are consistent with the demands of our study.  In addition, should the search find an especially exciting candidate, with false alarm rate of order once a year, it might be possible to involve space based observatories, such as SWIFT \cite{swift} or Chandra \cite{chan}.

\subsection{Image Processing and Data Analysis}

In addition to a well-defined, real time GW data analysis pipeline, this search requires an image processing data analysis pipeline to reduce images and identify any transients.  It is preferable to have this pipeline running in real-time as well.  Real-time transient identification would allow for a more refined observation schedule - one that follows up on found transients and ceases observations of targets where no transient is seen.

The recent interest in large field of view survey telescopes actively seeking transients has fueled development of image processing software (see for example \cite{trans} and \cite{trans2}).  To name a few, ROTSE, RAPTOR, Pan-STARRS, SkyMapper, and the Palomar Transient Factory all perform, or will perform, searches for transients \cite{rotse, raptor, PS, SM, PTF}.

 At the moment, we are working on an off line pipeline to process the images from the 2007 pilot studies (see section \ref{pilot}).  Our current approach is to reduce the images, and then to create photometry catalogs for each image.  The catalogs are compared in the time domain to search for objects exhibiting variability.

 There are a number of issues that are involved in defining such a pipeline.  We will have to find ways to classify variable objects - not all variability can be linked with highly energetic astrophysical processes.  In addition, it will be necessary to set thresholds to discern real variability from photometry uncertainty.

\section{Conclusions}

We have shown that the benefits of collaborations between GW observations and EM observations can be extended in scope by adding a search that begins with
interferometer triggers.  By using GW data to perform targeted transient searches, we take advantage of the nearly all-sky coverage of the LIGO-Virgo network.  From the GW end, such a search potentially extends the reach of ground-based interferometers by actively seeking EM verification of putative events.  From an astronomical perspective, this type of search has the possibility of triggering rapid observations of GRB orphan afterglows or other astro-physical transients.  In pilot studies, we have demonstrated the basic technique, and we plan to perform an extended search during the S6/VSR2 science run of the LIGO-Virgo network.  This study is extremely exciting, and represents a great step forward in the partnership between gravitational wave and electro-magnetic astronomy.

\ack

We would like to thank Shantanu Desai, Derek Fox, Ted Jacboson, and Richard O'Shaughnessy for helpful comments and discussions.
We would also like to thank Ehud Nakar and Avishay Gal-Yam for early discussions
which helped to inspire this project.  The authors gratefully acknowledge the support of the National Science
Foundation through grants PHY-06-53421 and PHY-04-57528, the University
of Maryland, Columbia University in the City of New York, and the
Carnegie Institution of Washington.  We also thank the staff and administration
of the MDM and Las Campanas observatories.

\end{document}